\documentclass[prl,superscriptaddress,showpacs,twocolumn]{revtex4-1}

\pdfoutput=1

\usepackage{graphicx}
\usepackage[
        pdfauthor={L. Steffen, M.P. da Silva, A. Fedorov, M. Baur,  and A. Wallraff},
        pdftitle={Experimental Monte Carlo Quantum Process Certification}
	]{hyperref}
\usepackage{subfigure}
\usepackage{amsmath}
\usepackage{amssymb}
\usepackage{dsfont}

\usepackage{color} 

\newcommand{\ket}[1]{| #1 \rangle}
\newcommand{\bra}[1]{\langle #1 |}
\newcommand{\ketbra}[2]{\ket{#1}\!\bra{#2}}

\begin{document}

\title{Experimental Monte Carlo Quantum Process Certification}

\author{L.~Steffen}
\affiliation{Department of Physics, ETH Zurich, CH-8093 Zurich, Switzerland}
\author{M.~P.~da~Silva}
\affiliation{Disruptive Information Processing Technologies Group, Raytheon BBN Technologies, 10 Moulton Street, Cambridge, MA 02138 USA}
\author{A.~Fedorov}
\author{M.~Baur}
\author{A.~Wallraff}
\affiliation{Department of Physics, ETH Zurich, CH-8093 Zurich, Switzerland}

\date{\today}
\begin{abstract}
Experimental implementations of quantum information processing have now reached a level of sophistication where quantum process tomography is impractical. The number of experimental settings as well as the computational cost of the data post-processing now translates to days of effort to characterize even experiments with as few as 8 qubits.
Recently a more practical approach to determine the fidelity of an experimental quantum process has been proposed, where the experimental data is compared directly to an ideal process using Monte Carlo sampling. Here we present an experimental implementation of this scheme in a circuit quantum electrodynamics setup
to determine the fidelity of two qubit gates, such as the \textsc{cphase} and the \textsc{cnot} gate, and three qubit gates, such as the Toffoli gate and two sequential \textsc{cphase} gates.
\end{abstract}


\maketitle
Quantum process tomography~\cite{Nielsen2000} is a widely used method to obtain a complete description of experimental implementations of gates or algorithms. With the ongoing experimental progress and growth in system size, quantum process tomography is already impractical and will soon become infeasible in state-of-the-art experiments, since the number of experimental settings as well as the computational cost of the post-processing increases exponentially with the number of qubits. Even the most recent tomography algorithms would need days of data post-processing in order to yield a process tomography estimate for as few as 8 qubits~\cite{Smolin2012}.

Experimentally, the determination of the process matrix $\chi$ for an $n$-qubit process involves the preparation of the qubits in $4^n$ different product states, where each qubit is prepared in one of the states $\ket{0}$, $\ket{0}+\ket{1}$, $\ket{0}-i\ket{1}$ or $\ket{1}$.
The quantum process under consideration is then allowed to act on each of these $n$-qubit initial states. Then the expectation values of $4^n$ linearly independent operations, typically chosen as all possible tensor products of Pauli operators, are measured. Overall, this results in $4^{2n}$ distinct expectation values to be measured. The process matrix $\chi_\mathrm{exp}$ can then be obtained by linear inversion.
However, experimental imperfections
and statistical fluctuations lead to unphysical results like a process matrix $\chi_\mathrm{exp}$ with negative eigenvalues or a trace unequal to one. To get a physical process matrix, one usually applies maximum-likelihood procedures  \cite{Hradil2004} which search for a physical process matrix $\chi_\mathrm{ML}$ that is most likely to have been implemented given the experimental observations. An estimate of the process fidelity~\cite{Schumacher1996,Gilchrist2005} between the implemented process and the ideal process is then calculated as
\begin{equation}
F(\chi_\mathrm{ideal},\chi_\mathrm{ML})=\mathrm{tr}\left[\chi_\mathrm{ideal}\chi_\mathrm{ML}\right],
\end{equation}
which in turn is related to the average output state fidelity $\overline F$ by $\overline F={(d F + 1) / (d + 1)}$ where $d$ is the dimension of the Hilbert space used to describe the states of the system~\cite{Horodecki1999} .

While quantum process tomography is useful to fully characterize a process, it has two major drawbacks for verifying processes by calculating their fidelity. First, it is inefficient, especially for a large number of qubits, since more information is acquired than is needed to calculate the process fidelity. Second, maximum-likelihood procedures can lead to estimates that imply greater confidence than can be supported from the data~\cite{Blume-Kohout2010}. Therefore, it is difficult to assign an error to the obtained fidelities. Instead, Monte Carlo process certification~\cite{Flammia2011, Silva2011} is an efficient approach for the determination of the fidelity of a quantum process. It does not rely on maximum-likelihood procedures and the number of measurements needed to obtain some desired accuracy depends only polynomially rather than exponentially on said accuracy, and not on the size of the system.

Here we present the implementation of Monte Carlo process certification on two- and three-qubit gates in a circuit QED system~\cite{Wallraff2004,Blais2004,Schoelkopf2008} with three transmon qubits~\cite{Koch2007} coupled to a superconducting waveguide resonator.
We give a detailed description of the protocol implemented with our 
setup and analyze the errors of the protocol. The obtained fidelities are then compared to fidelities obtained by quantum process tomography.

Monte Carlo process certification relies on the fact that an $n$-qubit process $\mathcal{E}$ can also be described by a $2n$-qubit density matrix $\hat{\rho}_\mathcal{E}=(\mathds{1}\otimes\mathcal{E})(\ketbra{\phi}{\phi})$, known as the Choi matrix~\cite{Jamiolkowski1972, Choi1975}, with $\ket{\phi}=\frac{1}{\sqrt{d}}\sum_{i=1}^d \ket{i}\otimes\ket{i}$ and $d=2^n$ the dimension of the state Hilbert space.  The fidelity of the experimentally realized process $\chi_\mathrm{exp}$ to the respective ideal unitary process $\chi_\mathrm{ideal}$ is identical to the fidelity of the corresponding Choi matrix
\begin{equation}
F(\chi_\mathrm{ideal},\chi_\mathrm{exp})= F(\hat{\rho}_\mathcal{E_\mathrm{ideal}},\hat{\rho}_\mathcal{E_\mathrm{exp}}) = \mathrm{tr}\left[\hat{\rho}_\mathcal{E_\mathrm{ideal}}\hat{\rho}_\mathcal{E_\mathrm{exp}}\right],
\end{equation}
where the last equality holds because the ideal process is unitary and thus the corresponding Choi matrix is pure. This expression can be re-written as
\begin{equation}
\label{eq:fid_final}
 F(\hat{\rho}_\mathcal{E_\mathrm{ideal}},\hat{\rho}_\mathcal{E_\mathrm{exp}}) =\sum_i \mathrm{Pr}(i)\frac{\sigma_i}{\rho_i},
\end{equation}
where $\rho_i=\mathrm{tr}\left[\hat{\rho}_\mathcal{E_\mathrm{ideal}}\hat{P}_i\right]$ and $\sigma_i=\mathrm{tr}\left[\hat{\rho}_\mathcal{E_\mathrm{exp}}\hat{P}_i\right]$. Here, $\hat{P}_i$ is an orthonormal Hermitian operator basis chosen as the $4^n$ tensor products of the Pauli matrices and the identity and the sum (\ref{eq:fid_final}) is taken over only the $i$ with $\rho_i\neq 0$. The distribution $\mathrm{Pr}(i)=\frac{\rho^2_i}{d}$ reflects the relevance of the expectation of an operator $\hat{P}_i$ for the fidelity calculation. By sampling randomly $N$ indices $i_1, i_2, \ldots i_N$ following the distribution $\mathrm{Pr}(i)$ one obtains an estimation of the fidelity $\frac{1}{N}\sum_{k=1}^N\frac{\sigma_{i_k}}{\rho_{i_k}}$ with an uncertainty that decreases as $\frac{1}{\sqrt{N}}$.

The straightforward implementation of Monte Carlo process certification as described above is rather impractical, since the measurement of a state $\hat{\rho}_\mathcal{E}$, representing the Choi matrix of the process $\mathcal{E}$,
would require $2n$ qubits for an $n$-qubit gate, as well as perfect storage of the $n$ ancillary qubits.

A more experimentally relevant approach is to prepare and measure $n$-qubit states only~\cite{Flammia2011, Silva2011}.
The key idea is that the effect of the measurement of the first half of the state $\ket{\phi}$, on which no gate is applied,
corresponds to a projection of the second half of the state $\ket{\phi}$ onto complex conjugates (in the computational basis) of eigenstates of the first half of the measurement operator. The measurement of $\hat{\rho}_\mathcal{E}$ with randomly chosen operators $\hat{A} \otimes \hat{B}$, where $\hat{A},\hat{B}$ are tensor products of $n$ Pauli matrices or identities, can be expressed as
\begin{align}
\label{eq:equiv}
\mathrm{tr}\left[(\hat{A}\otimes\hat{B})\hat{\rho}_\mathcal{E}\right]&=&\mathrm{tr}\left[(\hat{A}\otimes\hat{B})(\mathds{1}\otimes\mathcal{E})(\ketbra{\phi}{\phi})\right]\nonumber \\
&=&\frac{1}{d}\sum_{i=1}^d a_i\mathrm{tr}\left[\hat{B}~ \mathcal{E}(\ketbra{a_i}{a_i})\right].
\end{align}
Here $\ket{a_i}$ is the complex conjugate of the $i$\textsuperscript{th} eigenstate of the operator $\hat{A}$ with eigenvalue $a_i$. This final expression corresponds to the action of the process $\mathcal{E}$ on the state $\ket{a_i}$ followed by a measurement of the observable $\hat{B}$. The results for different input eigenstates are then summed up to obtain an estimate of $\mathrm{tr}\left[(\hat{A}\otimes\hat{B})\hat{\rho}_\mathcal{E}\right]$.

The implementation of the Monte Carlo process certification protocol was performed in a superconducting quantum processor consisting of three transmon qubits coupled to a coplanar waveguide resonator.  The sample used is the same as in Refs.~\cite{Baur2012,Fedorov2012}.

Our 3-qubit system is small enough that we can measure all relevant operators and do not need to resort to random sampling. This still allows for a significant saving in the number of measurements because many of the measurements required to perform process tomography are irrelevant for the fidelity estimate. In other words, we measure all operators that have a non-zero expectation value for the ideal gate, and calculate the accordingly weighted average to compute the gate fidelity.

The protocol requires the preparation of qubits in eigenstates of Pauli operators $\hat{A}$
and the measurement of Pauli operators $\hat{B}$.  The preparation of the qubit input states is straightforward by using amplitude and phase controlled coherent microwave pulses applied to the individual charge control lines. In our setup, the implementation of the measurement using joint dispersive readout~\cite{Filipp2009b} of all qubits is a more complex procedure. The measurement operator is
\begin{equation}
\hat{M}=\sum_{i_1,\ldots,i_n\in\{0,1\}}\alpha_{i_1,\ldots,i_n}\ketbra{i_1}{i_1}\otimes\ketbra{i_2}{i_2}\otimes\cdots\otimes\ketbra{i_n}{i_n},
\label{eq:opM}
\end{equation}
where $\ket{0}, \ket{1}$ are the computational basis states. The coefficients $\alpha_{i_1,\ldots,i_n}$ are obtained from measurements of the resonator transmission amplitude for each computational basis state \cite{Filipp2009b,Bianchetti2010,Bianchetti2009}. $\hat{M}$ expressed in terms of individual qubit identity and $\hat\sigma_z$ Pauli operators is
\begin{equation}
\hat{M}=\sum_{\hat{j}_1,\ldots,\hat{j}_n\in\{\mathds{1}
,\hat{\sigma}_z\}}\beta_{j_1,\ldots,j_n}
\hat{j}_1\otimes\hat{j}_2\otimes\cdots\otimes\hat{j}_n,
\end{equation}
with coefficients $\beta_{j_1,\ldots,j_n}$ calculated as combinations of the $\alpha_{i_1,\ldots,i_n}$.

In general the measurement operator has $2^n$ different elements. However, in Monte Carlo process certification for each input state the expectation value of only one specific element is needed. This element can be obtained by adding measurement outcomes with different signs of $\hat{\sigma}_z$ operators of different qubits, realized by $\pi$~pulses applied to the corresponding qubits just before the measurement. Since the first element $\mathds{1}\otimes\cdots\otimes\mathds{1}$ has always an expectation value of one, one needs to perform $2^{n-1}$ different measurement to extract a single operator $\hat{B}$.

As an example, the joint readout procedure of the operator  $\hat{\sigma}_y\otimes\hat{\sigma}_x$ for two qubits is presented in the following. The joint readout operator is
%
$
\hat{M}=\alpha_{00}\ketbra{0}{0} \otimes \ketbra{0}{0} + \alpha_{01}\ketbra{0}{0}\otimes\ketbra{1}{1}
 +\alpha_{10}\ketbra{1}{1}\otimes\ketbra{0}{0} +\alpha_{11} \ketbra{1}{1}\otimes\ketbra{1}{1},
$
%
which is equivalent to
$
\hat{M}=\beta_{00} \mathds{1} \otimes \mathds{1} + \beta_{01} \mathds{1}\otimes\hat{\sigma}_z
 +\beta_{10} \hat{\sigma}_z\otimes\mathds{1} +\beta_{11} \hat{\sigma}_z\otimes\hat{\sigma}_z.
$
%
The prefactors $\beta_{ij}$ are determined from measurements of the $\alpha_{ij}$ as described above.

To measure the given combination of Pauli operators, we rotate accordingly the measurement basis of the individual qubits. For the example above, we apply a $-\pi/2$ rotation around the $x$-axis to the first qubit and a $\pi/2$ rotation around the $y$-axis to the second qubit. The resulting measurement operator is
%
$
M=\beta_{00}\mathds{1} \otimes \mathds{1} + \beta_{01}\mathds{1}\otimes\hat{\sigma}_x
 +\beta_{10}\hat{\sigma}_y\otimes\mathds{1} +\beta_{11} \hat{\sigma}_y\otimes\hat{\sigma}_x.
$

To extract only the last term in the measurement operator, a second measurement with an additional $\pi$ pulse on both qubits is performed. This results in a measurement operator with two minus signs:
%
$
M=\beta_{00} \mathds{1} \otimes \mathds{1} - \beta_{01} \mathds{1}\otimes\hat{\sigma}_x -\beta_{10} \hat{\sigma}_y\otimes\mathds{1} +\beta_{11} \hat{\sigma}_y\otimes\hat{\sigma}_x.
$
%
Adding the measurement outcomes of the two experiments (for the same input state) gives the expectation value for the operator $2\left(\beta_{00} \mathds{1} \otimes \mathds{1} + \beta_{11} \hat{\sigma}_y\otimes\hat{\sigma}_x\right)$.
Since the expectation value for $\mathds{1} \otimes \mathds{1}$ is always equal to 1 and $\beta_{00}$ and $\beta_{11}$ are known, the expectation value of $\hat{\sigma}_y\otimes\hat{\sigma}_x$ can be extracted in this way.

\begin{figure*}[t]
\includegraphics[width=.99\textwidth]{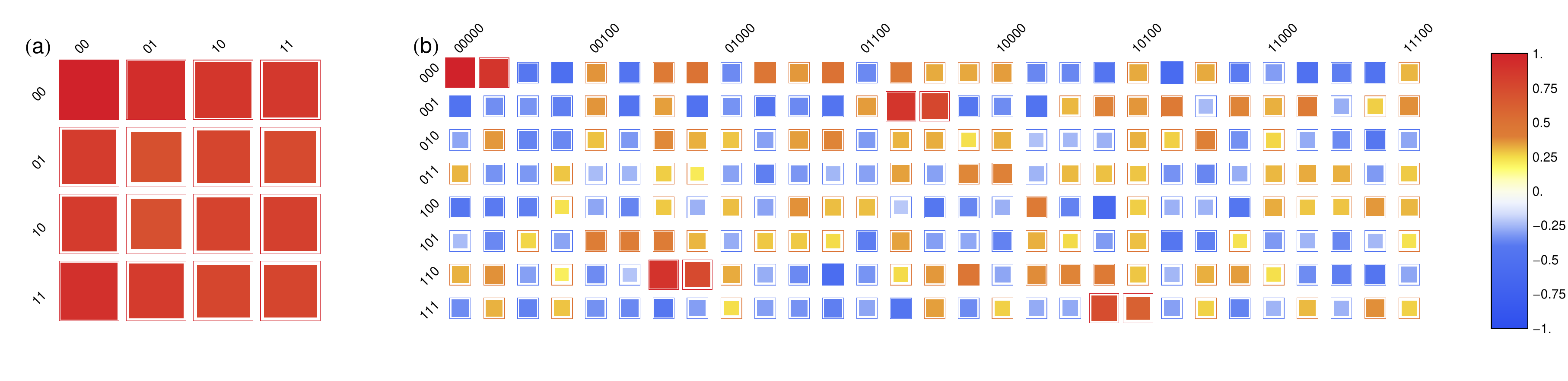}
\caption{Measured expectation values of all the relevant observables of the (a) \textsc{cnot} gate and (b) \textsc{Toffoli} gate Choi matrices. The thin border shows the ideal expected values, the colored squares are the estimated values. The $(00,00)$ and $(000,000)$ entries are the expectations of the identity, so they have sizes corresponding to absolute value 1 and the area of the other squares are adjusted proportionally. The column label corresponds to the most-significant digits of the binary expansion of the index of the observable, while the row label corresponds to the least significant digits (see supplementary information).}
\label{fig:hinton}
\end{figure*}

Hence, it is possible in our experiments to extract any expectation value of two-qubit Pauli operators from two measurements or three-qubit Pauli operators from four measurements, using the corresponding single qubit rotations. Having found the expectation values $\sigma_i = {\mathrm {tr}}\left[\hat{\rho}_{{\mathcal{E}}_{\rm{exp}}}\hat{P}_i\right]$, the fidelity can be directly calculated according to Eq.~(\ref{eq:fid_final}).

According to Eq.~(\ref{eq:equiv}), a measurement of one of the expectation values $\sigma_i$ consists of averaging measurement outcomes over different input states.
To achieve this, one can also perform a Monte Carlo sampling of which eigenvectors to prepare as input states. The weighting factor for the sampling is given by the absolute value of the eigenvalue.
Since in our experiments the system size is small but a high accuracy is desired, we measured all eigenstates.

The protocol has been tested on a 2-qubit  \textsc{cnot} and \textsc{cphase} gate \cite{Strauch2003, DiCarlo2009}, on a 3-qubit Toffoli gate \cite{Ralph2007, Fedorov2012}, and on the sequential application of two \textsc{cphase} gates on three qubits. The \textsc{cnot} and the \textsc{cphase} gates are particularly interesting for Monte Carlo process certification, since they map elements of the Pauli group to other elements of the Pauli group. Such gates are Clifford operations and their Choi matrices are stabilizer states~\cite{Gottesman1997, Gottesman1999} for which the number of relevant Pauli operators is minimal with uniform relevance distribution. For any stabilizer state $\hat\rho_{\mathcal{E}}$ there is a subgroup $S$ of the Pauli group with elements $\hat{S}_i$ such that the pure state corresponding to $\hat\rho_{\mathcal{E}_\mathrm{ideal}}$ is an eigenvector of all $ \hat{S}_i$ with eigenvalue $+1$. The expectation value of each operator in this stabilizer group is $+1$. Therefore, the relevance distribution $\mathrm{Pr}(i)=1/4^{n}$ is uniform for all $i \in\{1,\ldots, 4^n\}$. All other operators of the Pauli group have expectation value zero, and therefore have no impact on the estimation of the fidelity of a gate.

All experimentally realized gates have been characterized by calculation of their fidelity using Monte Carlo process certification ($F_\mathrm{MC}$), unconstrained tomography data ($F_\mathrm{tom}$), and tomography data constrained by maximum-likelihood estimation ($F_\mathrm{ML}$). 

The \textsc{cnot} gate, which changes the state of a target qubit if the control qubit is in the state $\ket{1}$, is described by a Choi matrix whose stabilizer group is generated by
\begin{equation}
\begin{array}{lcllll}
M_1&=&\hat{\sigma}_x & \mathds{1} & \hat{\sigma}_x & \hat{\sigma}_x,\\
M_2&=&\hat{\sigma}_z & \mathds{1} & \hat{\sigma}_z & \mathds{1},\\
M_3&=&\mathds{1} & \hat{\sigma}_x & \mathds{1} & \hat{\sigma}_x,\\
M_4&=&\mathds{1} & \hat{\sigma}_z & \hat{\sigma}_z &\hat{\sigma}_z.
\end{array}
\end{equation}
This indicates that, e.g. eigenstates of the $\hat{\sigma}_x \otimes \mathds{1}$ operator are mapped to eigenstates of the $\hat{\sigma}_x \otimes \hat{\sigma}_x$ operator by the \textsc{cnot} operation. A visualization of the expectation value of the 16 Pauli operators with non-vanishing relevance distribution is shown in Fig.~\ref{fig:hinton}(a). 
For the present gate, the total number of different measurement settings is $120$, since for each of the $15$ non-unity Pauli operators we prepare $4$ different input states and measure $2$ different operators (required by the joint readout). In contrast, the total number of different measurement settings for process tomography is $4^{(2\times 2)}=256$.

The \textsc{cphase} gate, which changes the phase of the $\ket{1}$ state of the target qubit by $\pi$ if the control qubit is in the state $\ket{1}$, has been characterized in a way similar to the \textsc{cnot} gate as these gates are locally equivalent.

A sequence of 2 \textsc{cphase} gates first acting on qubits 1 and 2, and then on qubits 2 and 3
was characterized as an example of a 3-qubit gate with a stabilizer state Choi matrix. This Choi matrix has $4^3 = 64$ Pauli operators with non-vanishing expectation value. For each of these operators we sample over 8 different eigenvectors by measuring $4$ different operator combinations (required by the joint readout), in total $2016$ different measurement settings, again without making use of random sampling.  In contrast, process tomography for any three-qubit gate requires $4^{2\times 3}=4096$ different measurement settings.

Our implementation of the Toffoli gate~\cite{Fedorov2012} was also characterized by Monte Carlo process certification and process tomography.
The Choi matrix of the Toffoli gate is not a stabilizer state. Therefore, the list of relevant Pauli operators has no group structure and the relevance distribution $\mathrm{Pr}(i)$ is not uniform. We find that there are $232$ Pauli operators with non-zero expectation value of $1$ or $\pm 0.5$ out of $4096$ possible ones. The total number of different relevant experimental settings is $231\times 8\times 4=7392$. 

Even without random sampling, the total number of measurements (including repeated measurements used for averaging) to achieve a smaller error is less for Monte Carlo process certification than for process tomography. For the Monte Carlo process estimation, we averaged each measurement setting $\sim330\,000$ times, resulting in a total number of $\sim 2.4~10^9$ measurements and an error of the fidelity of $0.5\%$, whereas for the process tomography we averaged each measurement setting for $\sim 790\,000$ times, resulting in a total number of $\sim 3.2~10^9$ measurements and an error of the fidelity of $3\%$. The measurement outcomes for the different operators are shown in Fig.~\ref{fig:hinton}(b).

All resulting fidelities are summarized in Table~\ref{tab:fidelities}. Errors are stated as $90\%$ confidence intervals. For Monte Carlo process estimation the error was calculated by Gaussian error propagation of the errors of the single measurements. For the error of the process tomography, the confidence interval of the distribution of fidelities was calculated based on a resampling of the measurement outcomes according to the inferred error statistics of the experiments. All the fidelities found with Monte Carlo process certification have tighter error bars than the fidelities obtained from process tomography. This is mainly due to the fact that the postprocessing for the Monte Carlo certification only consists of averaging the relevant measured values whereas full process tomography must impose collective physical constraints on the entire data set, and errors on the irrelevant observables can only add to the errors relating to the relevant observables.

\begin{table}[b]
\begin{tabular}{l|c|c|c}
Gate&$F_\mathrm{MC}$&$F_\mathrm{tom}$&$F_\mathrm{ML}$\\ \hline
\textsc{cnot}&$81.7 \pm 2.1 \%$&$80 \pm 3\%$&$79 \pm 3 \%$\\
\textsc{cphase}&$86.6 \pm3.0 \%$&$86 \pm 4\%$&$83 \pm 4 \%$\\
2 \textsc{cphase}s&$65.0 \pm 0.8 \%$&$67 \pm 5 \%$&$67 \pm 5 \%$\\
Toffoli&$68.5 \pm 0.5 \%$&$ 70 \pm 3\%$&$69 \pm 3 \%$\\
\end{tabular}
\caption{Fidelities obtained by Monte Carlo process certification ($F_\mathrm{MC}$) compared to the values obtained with process tomography ($F_\mathrm{tom}$) and subsequent application of a maximum likelihood algorithm ($F_\mathrm{ML}$).}
\label{tab:fidelities}
\end{table}

\begin{figure}
\includegraphics[width=0.48\textwidth]{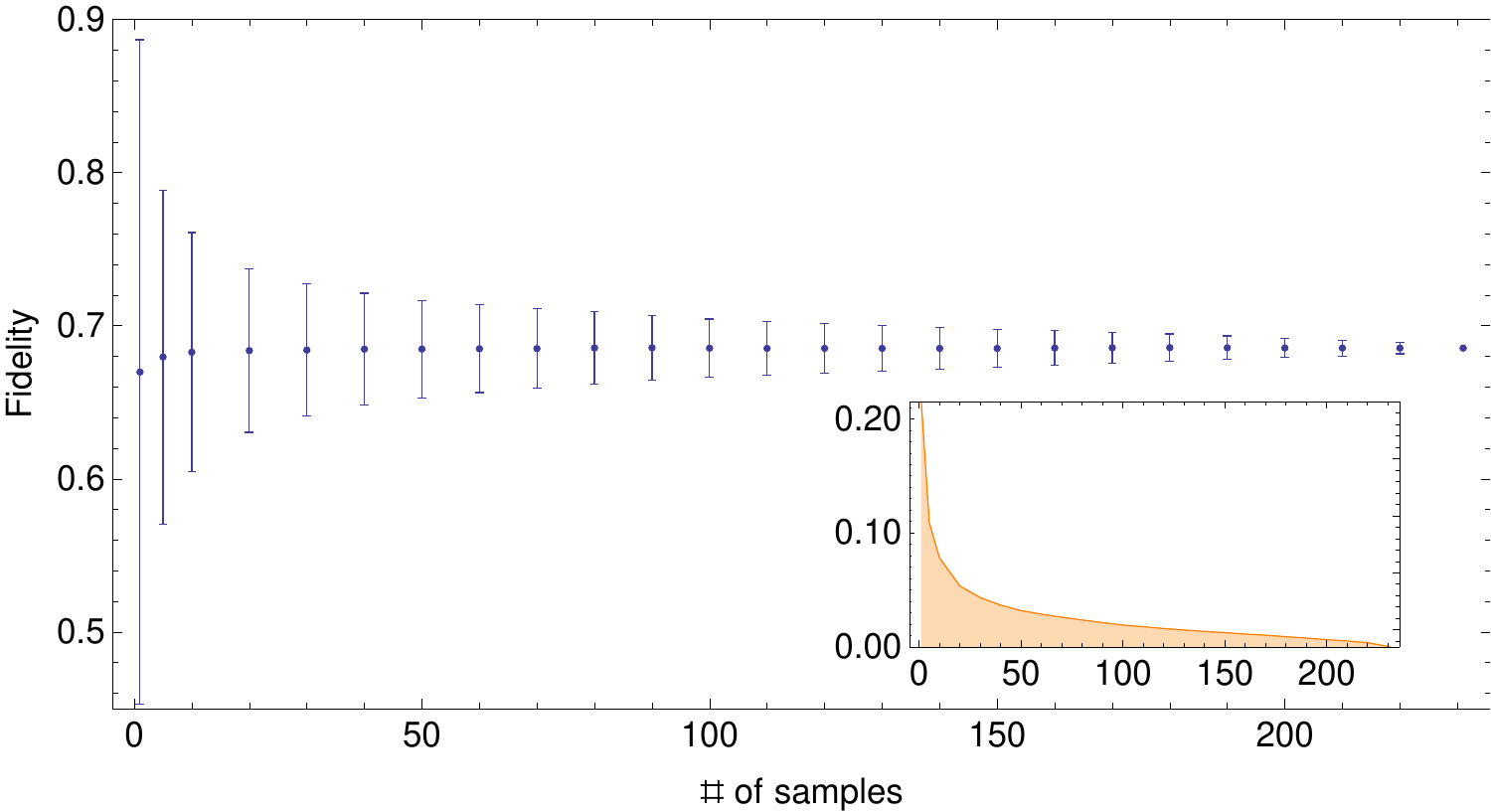}
\caption{Mean of the estimated average output fidelity of a Toffoli gate as a function of the number of sampled observables. The error bars correspond to the $90 \%$ confidence intervals, which in turns gives an estimate of the additional error due to the non-exhaustive sampling of relevant observables. The inset gives the half-width of the $90 \%$ confidence intervals for the corresponding number of samples.}
\label{fig:toffoli_mc_errors}
\end{figure}

As described earlier, the significant advantage of Monte Carlo process estimation is that one can estimate the fidelity of a process also without sampling over all relevant Pauli operators, on the expense of a higher uncertainty.  If all relevant Pauli operators have been measured like in our experiments, the only error in the fidelity is due to the experimental uncertainty in the estimation of the different expectation values. In the case that an incomplete set of Pauli operators is sampled, there is an additional error.  An asymptotic bound for this error is calculated in the supplementary material of Ref.~\cite{Silva2011}, and it is shown that these bounds scale polynomially with the number of measured samples. However, the bounds are not tight and therefore too pessimistic to be used in the calculation of error bars.
The error in the fidelity estimate when performing non-exhaustive sampling of the Pauli operators can be obtained by non-parametric resampling methods such as {\em bootstrapping}~\cite{Efron1981}.  However, given that we have measured all the relevant Pauli operators for each of the gates we characterized, we can simply gather statistics for estimates with non-exhaustive sampling.  The corresponding data for the Toffoli gate is shown in Fig.~\ref{fig:toffoli_mc_errors}. For our data one gets e.\,g.~an additional error of $2 \%$ if one only samples $100$ Pauli operators or an additional error of $3.2 \%$ for sampling only $50$ Pauli operators. This illustrates that Monte Carlo sampling leads to significant reduction in the numbers of measurements required to achive a given error bound on the fidelities.

In conclusion we showed how Monte Carlo process certification can be implemented experimentally in a system with three qubits and joint readout. This scheme is generic and readily applicable to any qubit system. We characterized the fidelity of two 2-qubit- and two 3-qubit gates. All estimates of the gate fidelity for each of the four gates are consistent, although Monte Carlo process certification gives more accurate estimates of the fidelity using fewer measurements. This shows that Monte Carlo process certification can be used as an independent proof of the fidelity.

This work was supported by the Swiss National Science Foundation (SNF), the EU IP SOLID, and ETH Zurich.

\bibliographystyle{apsrev4-1}
\bibliography{mc_process_verification}

\end{document}


\renewcommand{\thefigure}{S\arabic{figure}}
\renewcommand{\thetable}{S\arabic{table}}
\renewcommand{\theequation}{S\arabic{equation}}

\title{Experimental Monte Carlo Quantum Process Certification : Supplemental Material}

\author{L.~Steffen}
\affiliation{Department of Physics, ETH Zurich, CH-8093 Zurich, Switzerland}
\author{M.~P.~da~Silva}
\affiliation{Disruptive Information Processing Technologies Group, Raytheon BBN Technologies, 10 Moulton Street, Cambridge, MA 02138 USA}
\author{A.~Fedorov}
\author{M.~Baur}
\author{A.~Wallraff}
\affiliation{Department of Physics, ETH Zurich, CH-8093 Zurich, Switzerland}

\date{\today}

\maketitle
In this supplemental material we summarize the experimental protocol that is used and give lists of all relevant operators that were used for the measurements described in the main letter.
\section{Experimental protocol}
The following recipe summarizes the experimental protocol described in the letter:
\begin{enumerate}
 \item For a given ideal process $\mathcal{E}$ calculate the Choi matrix $\hat{\rho}_{\mathcal{E}_\mathrm{ideal}}$.
 \item Find Pauli operators $\hat{P}_i=\hat{A}_i\otimes\hat{B}_i$ which have non-vanishing expectation values $\rho_i=\mathrm{tr}\hat{\rho}_{\mathcal{E}_\mathrm{ideal}}\hat{P}_i\neq 0$.
 \item For every Pauli operator $\hat{P}_i$ find the $\ket{a_{i_1}},\ket{a_{i_2}},\ldots,\ket{a_{i_d}}$ (complex conjugates of the eigenvectors of $\hat{A}_i$), and do the following:
  \begin{enumerate}
    \item Prepare the gate inputs in the state $\ket{a_{i_j}}$.
    \item Apply the process $\mathcal{E}_\mathrm{exp}$ to the state prepared in (a).
    \item Measure the expectation $\hat{B}_i$ given the output of (b).
    \item Compute the weighted average of the expectation of $\hat{B}_i$ over all the complex conjugates of the eigenstates, using the corresponding eigenvalues as the weight. This gives an estimate of $\sigma_i$ in the computation of the gate fidelity.
  \end{enumerate}
 \item Calculate the average of the estimated outcomes $\sigma_i$ weighted with the relevance distribution $\mathrm{Pr}(i)$ according to Eq.~(3) of the main publication to get the fidelity. 
\end{enumerate}

\section{Relevant operators for the cnot gate}
The stabilizer group that describes the Choi matrix of the \textsc{cnot}-gate has generators
\begin{equation}
\begin{array}{lcllll}
M_1&=&\hat{\sigma}_x & \mathds{1} & \hat{\sigma}_x & \hat{\sigma}_x,\\
M_2&=&\hat{\sigma}_z & \mathds{1} & \hat{\sigma}_z & \mathds{1},\\
M_3&=&\mathds{1} & \hat{\sigma}_x & \mathds{1} & \hat{\sigma}_x,\\
M_4&=&\mathds{1} & \hat{\sigma}_z & \hat{\sigma}_z &\hat{\sigma}_z.
\end{array}
\end{equation}

The relevant Pauli operators are the products of the elements of each subset of these generators. A complete list of the relevant operators is shown in Tab.~\ref{tab:cnotops}. Each block of four operators corresponds to a row of the plot in Fig.~1 (a) of the main publication.

\begin{table}
\begin{math}
\begin{array}{lllll}
&\mathds{1}&\mathds{1}&\mathds{1}&\mathds{1}\\
&\mathds{1}&\hat{\sigma}_x&\mathds{1}&\hat{\sigma}_x\\
-&\mathds{1}&\hat{\sigma}_y&\hat{\sigma}_z&\hat{\sigma}_y\\
&\mathds{1}&\hat{\sigma}_z&\hat{\sigma}_z&\hat{\sigma}_z\\
\hline
&\hat{\sigma}_x&\mathds{1}&\hat{\sigma}_x&\hat{\sigma}_x\\
&\hat{\sigma}_x&\hat{\sigma}_x&\hat{\sigma}_x&\mathds{1}\\
-&\hat{\sigma}_x&\hat{\sigma}_y&\hat{\sigma}_y&\hat{\sigma}_z\\
-&\hat{\sigma}_x&\hat{\sigma}_z&\hat{\sigma}_y&\hat{\sigma}_y\\
\hline
-&\hat{\sigma}_y&\mathds{1}&\hat{\sigma}_y&\hat{\sigma}_x\\
-&\hat{\sigma}_y&\hat{\sigma}_x&\hat{\sigma}_y&\mathds{1}\\
-&\hat{\sigma}_y&\hat{\sigma}_y&\hat{\sigma}_x&\hat{\sigma}_z\\
-&\hat{\sigma}_y&\hat{\sigma}_z&\hat{\sigma}_x&\hat{\sigma}_y\\
\hline
&\hat{\sigma}_z&\mathds{1}&\hat{\sigma}_z&\mathds{1}\\
&\hat{\sigma}_z&\hat{\sigma}_x&\hat{\sigma}_z&\hat{\sigma}_x\\
-&\hat{\sigma}_z&\hat{\sigma}_y&\mathds{1}&\hat{\sigma}_y\\
&\hat{\sigma}_z&\hat{\sigma}_z&\mathds{1}&\hat{\sigma}_z
\end{array}
\end{math}
\caption{All relevant Pauli operators for the verification of a \textsc{cnot} gate.}
\label{tab:cnotops}
\end{table}

\section{Relevant operators for the cphase gate}
The stabilizer group that describes the Choi matrix of the \textsc{cphase}-gate has generators
\begin{equation}
\begin{array}{lcllll}
M_1&=&\hat{\sigma}_x & \mathds{1} & \hat{\sigma}_x & \hat{\sigma}_z,\\
M_2&=&\hat{\sigma}_z & \mathds{1} & \hat{\sigma}_z & \mathds{1},\\
M_3&=&\mathds{1} & \hat{\sigma}_x & \hat{\sigma}_z & \hat{\sigma}_x,\\
M_4&=&\mathds{1} & \hat{\sigma}_z & \mathds{1} &\hat{\sigma}_z.
\end{array}
\end{equation}

The relevant Pauli operators are found in the same way as for the \textsc{cnot} gate and are listed in Tab.~\ref{tab:cphaseops}.

\begin{table}
\begin{math}
\begin{array}{lllll}
&\mathds{1}&\mathds{1}&\mathds{1}&\mathds{1}\\
&\mathds{1}&\hat{\sigma}_x&\hat{\sigma}_z&\hat{\sigma}_x\\
-&\mathds{1}&\hat{\sigma}_y&\hat{\sigma}_z&\hat{\sigma}_y\\
&\mathds{1}&\hat{\sigma}_z&\mathds{1}&\hat{\sigma}_z\\
\hline
&\hat{\sigma}_x&\mathds{1}&\hat{\sigma}_x&\hat{\sigma}_z\\
&\hat{\sigma}_x&\hat{\sigma}_x&\hat{\sigma}_y&\hat{\sigma}_y\\
&\hat{\sigma}_x&\hat{\sigma}_y&\hat{\sigma}_y&\hat{\sigma}_x\\
&\hat{\sigma}_x&\hat{\sigma}_z&\hat{\sigma}_x&\mathds{1}\\
\hline
-&\hat{\sigma}_y&\mathds{1}&\hat{\sigma}_y&\hat{\sigma}_z\\
&\hat{\sigma}_y&\hat{\sigma}_x&\hat{\sigma}_x&\hat{\sigma}_y\\
&\hat{\sigma}_y&\hat{\sigma}_y&\hat{\sigma}_x&\hat{\sigma}_x\\
-&\hat{\sigma}_y&\hat{\sigma}_z&\hat{\sigma}_y&\mathds{1}\\
\hline
&\hat{\sigma}_z&\mathds{1}&\hat{\sigma}_z&\mathds{1}\\
&\hat{\sigma}_z&\hat{\sigma}_x&\mathds{1}&\hat{\sigma}_x\\
-&\hat{\sigma}_z&\hat{\sigma}_y&\mathds{1}&\hat{\sigma}_y\\
&\hat{\sigma}_z&\hat{\sigma}_z&\hat{\sigma}_z&\hat{\sigma}_z
\end{array}
\end{math}
\caption{All relevant Pauli operators for the verification of a \textsc{cphase} gate.}
\label{tab:cphaseops}
\end{table}

\section{Relevant operators for two sequential cphase gates}
A sequence of two \textsc{cphase}-gates overlapping on the middle qubit has stabilizer group generators
\begin{equation}
\begin{array}{lcllllll}
M_1&=&\hat{\sigma}_x & \mathds{1} & \mathds{1} & \hat{\sigma}_x & \hat{\sigma}_z& \mathds{1} ,\\
M_2&=&\hat{\sigma}_z & \mathds{1} & \mathds{1}  & \hat{\sigma}_z & \mathds{1}& \mathds{1} ,\\
M_3&=&\mathds{1} & \hat{\sigma}_x & \mathds{1}  & \hat{\sigma}_z & \hat{\sigma}_x & \hat{\sigma}_z,\\
M_4&=&\mathds{1} & \hat{\sigma}_z & \mathds{1} & \mathds{1} &\hat{\sigma}_z & \mathds{1},\\
M_5&=&\mathds{1}  & \mathds{1} & \hat{\sigma}_x& \mathds{1} &\hat{\sigma}_z&\hat{\sigma}_x,\\
M_6&=&\mathds{1}  & \mathds{1} & \hat{\sigma}_z& \mathds{1}& \mathds{1}&\hat{\sigma}_z.\\
\end{array}
\end{equation}

The relevant Pauli operators found from these generators are listed in Tab.~\ref{tab:cphase2}.

\begin{table*}
\begin{math}
\begin{array}{llllllllllllllllllllllllllllllll}
&\mathds{1}&\mathds{1}&\mathds{1}&\mathds{1}&\mathds{1}&\mathds{1}&\hspace{0.7cm}&&\hat{\sigma}_x&\mathds{1}&\mathds{1}&\hat{\sigma}_x&\hat{\sigma}_z&\mathds{1}&\hspace{0.7cm}&-&\hat{\sigma}_y&\mathds{1}&\mathds{1}&\hat{\sigma}_y&\hat{\sigma}_z&\mathds{1}&\hspace{0.7cm}&&\hat{\sigma}_z&\mathds{1}&\mathds{1}&\hat{\sigma}_z&\mathds{1}&\mathds{1}\\
&\mathds{1}&\mathds{1}&\hat{\sigma}_x&\mathds{1}&\hat{\sigma}_z&\hat{\sigma}_x&\hspace{0.7cm}&&\hat{\sigma}_x&\mathds{1}&\hat{\sigma}_x&\hat{\sigma}_x&\mathds{1}&\hat{\sigma}_x&\hspace{0.7cm}&-&\hat{\sigma}_y&\mathds{1}&\hat{\sigma}_x&\hat{\sigma}_y&\mathds{1}&\hat{\sigma}_x&\hspace{0.7cm}&&\hat{\sigma}_z&\mathds{1}&\hat{\sigma}_x&\hat{\sigma}_z&\hat{\sigma}_z&\hat{\sigma}_x\\
-&\mathds{1}&\mathds{1}&\hat{\sigma}_y&\mathds{1}&\hat{\sigma}_z&\hat{\sigma}_y&\hspace{0.7cm}&-&\hat{\sigma}_x&\mathds{1}&\hat{\sigma}_y&\hat{\sigma}_x&\mathds{1}&\hat{\sigma}_y&\hspace{0.7cm}&&\hat{\sigma}_y&\mathds{1}&\hat{\sigma}_y&\hat{\sigma}_y&\mathds{1}&\hat{\sigma}_y&\hspace{0.7cm}&-&\hat{\sigma}_z&\mathds{1}&\hat{\sigma}_y&\hat{\sigma}_z&\hat{\sigma}_z&\hat{\sigma}_y\\
&\mathds{1}&\mathds{1}&\hat{\sigma}_z&\mathds{1}&\mathds{1}&\hat{\sigma}_z&\hspace{0.7cm}&&\hat{\sigma}_x&\mathds{1}&\hat{\sigma}_z&\hat{\sigma}_x&\hat{\sigma}_z&\hat{\sigma}_z&\hspace{0.7cm}&-&\hat{\sigma}_y&\mathds{1}&\hat{\sigma}_z&\hat{\sigma}_y&\hat{\sigma}_z&\hat{\sigma}_z&\hspace{0.7cm}&&\hat{\sigma}_z&\mathds{1}&\hat{\sigma}_z&\hat{\sigma}_z&\mathds{1}&\hat{\sigma}_z\\
\cline{1-7}\cline{9-15}\cline{17-23}\cline{25-31}
&\mathds{1}&\hat{\sigma}_x&\mathds{1}&\hat{\sigma}_z&\hat{\sigma}_x&\hat{\sigma}_z&\hspace{0.7cm}&&\hat{\sigma}_x&\hat{\sigma}_x&\mathds{1}&\hat{\sigma}_y&\hat{\sigma}_y&\hat{\sigma}_z&\hspace{0.7cm}&&\hat{\sigma}_y&\hat{\sigma}_x&\mathds{1}&\hat{\sigma}_x&\hat{\sigma}_y&\hat{\sigma}_z&\hspace{0.7cm}&&\hat{\sigma}_z&\hat{\sigma}_x&\mathds{1}&\mathds{1}&\hat{\sigma}_x&\hat{\sigma}_z\\
&\mathds{1}&\hat{\sigma}_x&\hat{\sigma}_x&\hat{\sigma}_z&\hat{\sigma}_y&\hat{\sigma}_y&\hspace{0.7cm}&-&\hat{\sigma}_x&\hat{\sigma}_x&\hat{\sigma}_x&\hat{\sigma}_y&\hat{\sigma}_x&\hat{\sigma}_y&\hspace{0.7cm}&-&\hat{\sigma}_y&\hat{\sigma}_x&\hat{\sigma}_x&\hat{\sigma}_x&\hat{\sigma}_x&\hat{\sigma}_y&\hspace{0.7cm}&&\hat{\sigma}_z&\hat{\sigma}_x&\hat{\sigma}_x&\mathds{1}&\hat{\sigma}_y&\hat{\sigma}_y\\
&\mathds{1}&\hat{\sigma}_x&\hat{\sigma}_y&\hat{\sigma}_z&\hat{\sigma}_y&\hat{\sigma}_x&\hspace{0.7cm}&-&\hat{\sigma}_x&\hat{\sigma}_x&\hat{\sigma}_y&\hat{\sigma}_y&\hat{\sigma}_x&\hat{\sigma}_x&\hspace{0.7cm}&-&\hat{\sigma}_y&\hat{\sigma}_x&\hat{\sigma}_y&\hat{\sigma}_x&\hat{\sigma}_x&\hat{\sigma}_x&\hspace{0.7cm}&&\hat{\sigma}_z&\hat{\sigma}_x&\hat{\sigma}_y&\mathds{1}&\hat{\sigma}_y&\hat{\sigma}_x\\
&\mathds{1}&\hat{\sigma}_x&\hat{\sigma}_z&\hat{\sigma}_z&\hat{\sigma}_x&\mathds{1}&\hspace{0.7cm}&&\hat{\sigma}_x&\hat{\sigma}_x&\hat{\sigma}_z&\hat{\sigma}_y&\hat{\sigma}_y&\mathds{1}&\hspace{0.7cm}&&\hat{\sigma}_y&\hat{\sigma}_x&\hat{\sigma}_z&\hat{\sigma}_x&\hat{\sigma}_y&\mathds{1}&\hspace{0.7cm}&&\hat{\sigma}_z&\hat{\sigma}_x&\hat{\sigma}_z&\mathds{1}&\hat{\sigma}_x&\mathds{1}\\
\cline{1-7}\cline{9-15}\cline{17-23}\cline{25-31}
-&\mathds{1}&\hat{\sigma}_y&\mathds{1}&\hat{\sigma}_z&\hat{\sigma}_y&\hat{\sigma}_z&\hspace{0.7cm}&&\hat{\sigma}_x&\hat{\sigma}_y&\mathds{1}&\hat{\sigma}_y&\hat{\sigma}_x&\hat{\sigma}_z&\hspace{0.7cm}&&\hat{\sigma}_y&\hat{\sigma}_y&\mathds{1}&\hat{\sigma}_x&\hat{\sigma}_x&\hat{\sigma}_z&\hspace{0.7cm}&-&\hat{\sigma}_z&\hat{\sigma}_y&\mathds{1}&\mathds{1}&\hat{\sigma}_y&\hat{\sigma}_z\\
&\mathds{1}&\hat{\sigma}_y&\hat{\sigma}_x&\hat{\sigma}_z&\hat{\sigma}_x&\hat{\sigma}_y&\hspace{0.7cm}&&\hat{\sigma}_x&\hat{\sigma}_y&\hat{\sigma}_x&\hat{\sigma}_y&\hat{\sigma}_y&\hat{\sigma}_y&\hspace{0.7cm}&&\hat{\sigma}_y&\hat{\sigma}_y&\hat{\sigma}_x&\hat{\sigma}_x&\hat{\sigma}_y&\hat{\sigma}_y&\hspace{0.7cm}&&\hat{\sigma}_z&\hat{\sigma}_y&\hat{\sigma}_x&\mathds{1}&\hat{\sigma}_x&\hat{\sigma}_y\\
&\mathds{1}&\hat{\sigma}_y&\hat{\sigma}_y&\hat{\sigma}_z&\hat{\sigma}_x&\hat{\sigma}_x&\hspace{0.7cm}&&\hat{\sigma}_x&\hat{\sigma}_y&\hat{\sigma}_y&\hat{\sigma}_y&\hat{\sigma}_y&\hat{\sigma}_x&\hspace{0.7cm}&&\hat{\sigma}_y&\hat{\sigma}_y&\hat{\sigma}_y&\hat{\sigma}_x&\hat{\sigma}_y&\hat{\sigma}_x&\hspace{0.7cm}&&\hat{\sigma}_z&\hat{\sigma}_y&\hat{\sigma}_y&\mathds{1}&\hat{\sigma}_x&\hat{\sigma}_x\\
-&\mathds{1}&\hat{\sigma}_y&\hat{\sigma}_z&\hat{\sigma}_z&\hat{\sigma}_y&\mathds{1}&\hspace{0.7cm}&&\hat{\sigma}_x&\hat{\sigma}_y&\hat{\sigma}_z&\hat{\sigma}_y&\hat{\sigma}_x&\mathds{1}&\hspace{0.7cm}&&\hat{\sigma}_y&\hat{\sigma}_y&\hat{\sigma}_z&\hat{\sigma}_x&\hat{\sigma}_x&\mathds{1}&\hspace{0.7cm}&-&\hat{\sigma}_z&\hat{\sigma}_y&\hat{\sigma}_z&\mathds{1}&\hat{\sigma}_y&\mathds{1}\\
\cline{1-7}\cline{9-15}\cline{17-23}\cline{25-31}
&\mathds{1}&\hat{\sigma}_z&\mathds{1}&\mathds{1}&\hat{\sigma}_z&\mathds{1}&\hspace{0.7cm}&&\hat{\sigma}_x&\hat{\sigma}_z&\mathds{1}&\hat{\sigma}_x&\mathds{1}&\mathds{1}&\hspace{0.7cm}&-&\hat{\sigma}_y&\hat{\sigma}_z&\mathds{1}&\hat{\sigma}_y&\mathds{1}&\mathds{1}&\hspace{0.7cm}&&\hat{\sigma}_z&\hat{\sigma}_z&\mathds{1}&\hat{\sigma}_z&\hat{\sigma}_z&\mathds{1}\\
&\mathds{1}&\hat{\sigma}_z&\hat{\sigma}_x&\mathds{1}&\mathds{1}&\hat{\sigma}_x&\hspace{0.7cm}&&\hat{\sigma}_x&\hat{\sigma}_z&\hat{\sigma}_x&\hat{\sigma}_x&\hat{\sigma}_z&\hat{\sigma}_x&\hspace{0.7cm}&-&\hat{\sigma}_y&\hat{\sigma}_z&\hat{\sigma}_x&\hat{\sigma}_y&\hat{\sigma}_z&\hat{\sigma}_x&\hspace{0.7cm}&&\hat{\sigma}_z&\hat{\sigma}_z&\hat{\sigma}_x&\hat{\sigma}_z&\mathds{1}&\hat{\sigma}_x\\
-&\mathds{1}&\hat{\sigma}_z&\hat{\sigma}_y&\mathds{1}&\mathds{1}&\hat{\sigma}_y&\hspace{0.7cm}&-&\hat{\sigma}_x&\hat{\sigma}_z&\hat{\sigma}_y&\hat{\sigma}_x&\hat{\sigma}_z&\hat{\sigma}_y&\hspace{0.7cm}&&\hat{\sigma}_y&\hat{\sigma}_z&\hat{\sigma}_y&\hat{\sigma}_y&\hat{\sigma}_z&\hat{\sigma}_y&\hspace{0.7cm}&-&\hat{\sigma}_z&\hat{\sigma}_z&\hat{\sigma}_y&\hat{\sigma}_z&\mathds{1}&\hat{\sigma}_y\\
&\mathds{1}&\hat{\sigma}_z&\hat{\sigma}_z&\mathds{1}&\hat{\sigma}_z&\hat{\sigma}_z&\hspace{0.7cm}&&\hat{\sigma}_x&\hat{\sigma}_z&\hat{\sigma}_z&\hat{\sigma}_x&\mathds{1}&\hat{\sigma}_z&\hspace{0.7cm}&-&\hat{\sigma}_y&\hat{\sigma}_z&\hat{\sigma}_z&\hat{\sigma}_y&\mathds{1}&\hat{\sigma}_z&\hspace{0.7cm}&&\hat{\sigma}_z&\hat{\sigma}_z&\hat{\sigma}_z&\hat{\sigma}_z&\hat{\sigma}_z&\hat{\sigma}_z\\
\end{array}
\end{math}
\caption{All relevant Pauli operators for the verification of two sequential \textsc{cphase} gates}
\label{tab:cphase2}
\end{table*}

\section{Relevant operators for the Toffoli gate}
Since the Choi matrix of the Toffoli gate is not a stabilizer state, the list of relevant Pauli operators has no group structure and the relevance distribution $\mathrm{Pr}(i)$ is not uniform. By calculating explicitly the expectation value for all 4096 possible Pauli operators, we find only 232 which are non-zero. The corresponding operators are listed in Tabs.~\ref{tab:toffoli1} and \ref{tab:toffoli2}, labeld consistently with the measured values shown in Fig.~1(b) of the main letter.

\begin{table*}
\begin{math}
\begin{array}{llllllllllllllllllllllllllllll}
&&&\multicolumn{6}{c}{000}&&\multicolumn{6}{c}{001}&&\multicolumn{6}{c}{010}&&\multicolumn{6}{c}{011}\\\\\textstyle{00000}&\hspace{0.1cm}&&\mathds{1}&\mathds{1}&\mathds{1}&\mathds{1}&\mathds{1}&\mathds{1}&\hspace{0.7cm}&\mathds{1}&\hat{\sigma}_y&\mathds{1}&\hat{\sigma}_z&\hat{\sigma}_y&\hat{\sigma}_x&\hspace{0.7cm}&\hat{\sigma}_x&\mathds{1}&\hat{\sigma}_x&\hat{\sigma}_x&\hat{\sigma}_z&\mathds{1}&\hspace{0.7cm}&\hat{\sigma}_x&\hat{\sigma}_y&\mathds{1}&\hat{\sigma}_y&\hat{\sigma}_x&\hat{\sigma}_x\\
\textstyle{}&\hspace{0.1cm}&&\mathds{1}&\mathds{1}&\hat{\sigma}_x&\mathds{1}&\mathds{1}&\hat{\sigma}_x&\hspace{0.7cm}&\mathds{1}&\hat{\sigma}_y&\hat{\sigma}_x&\mathds{1}&\hat{\sigma}_y&\mathds{1}&\hspace{0.7cm}&\hat{\sigma}_x&\mathds{1}&\hat{\sigma}_x&\hat{\sigma}_x&\hat{\sigma}_z&\hat{\sigma}_x&\hspace{0.7cm}&\hat{\sigma}_x&\hat{\sigma}_y&\hat{\sigma}_x&\hat{\sigma}_x&\hat{\sigma}_y&\mathds{1}\\
\textstyle{}&\hspace{0.1cm}&&\mathds{1}&\mathds{1}&\hat{\sigma}_y&\mathds{1}&\mathds{1}&\hat{\sigma}_y&\hspace{0.7cm}&\mathds{1}&\hat{\sigma}_y&\hat{\sigma}_x&\mathds{1}&\hat{\sigma}_y&\hat{\sigma}_x&\hspace{0.7cm}&\hat{\sigma}_x&\mathds{1}&\hat{\sigma}_y&\hat{\sigma}_x&\mathds{1}&\hat{\sigma}_y&\hspace{0.7cm}&\hat{\sigma}_x&\hat{\sigma}_y&\hat{\sigma}_x&\hat{\sigma}_x&\hat{\sigma}_y&\hat{\sigma}_x\\
\textstyle{}&\hspace{0.1cm}&&\mathds{1}&\mathds{1}&\hat{\sigma}_y&\mathds{1}&\hat{\sigma}_z&\hat{\sigma}_y&\hspace{0.7cm}&\mathds{1}&\hat{\sigma}_y&\hat{\sigma}_x&\hat{\sigma}_z&\hat{\sigma}_y&\mathds{1}&\hspace{0.7cm}&\hat{\sigma}_x&\mathds{1}&\hat{\sigma}_y&\hat{\sigma}_x&\hat{\sigma}_z&\hat{\sigma}_y&\hspace{0.7cm}&\hat{\sigma}_x&\hat{\sigma}_y&\hat{\sigma}_x&\hat{\sigma}_y&\hat{\sigma}_x&\mathds{1}\\
\textstyle{00100}&\hspace{0.1cm}&&\mathds{1}&\mathds{1}&\hat{\sigma}_y&\hat{\sigma}_z&\mathds{1}&\hat{\sigma}_y&\hspace{0.7cm}&\mathds{1}&\hat{\sigma}_y&\hat{\sigma}_x&\hat{\sigma}_z&\hat{\sigma}_y&\hat{\sigma}_x&\hspace{0.7cm}&\hat{\sigma}_x&\mathds{1}&\hat{\sigma}_y&\hat{\sigma}_y&\mathds{1}&\hat{\sigma}_z&\hspace{0.7cm}&\hat{\sigma}_x&\hat{\sigma}_y&\hat{\sigma}_x&\hat{\sigma}_y&\hat{\sigma}_x&\hat{\sigma}_x\\
\textstyle{}&\hspace{0.1cm}&&\mathds{1}&\mathds{1}&\hat{\sigma}_y&\hat{\sigma}_z&\hat{\sigma}_z&\hat{\sigma}_y&\hspace{0.7cm}&\mathds{1}&\hat{\sigma}_y&\hat{\sigma}_y&\mathds{1}&\hat{\sigma}_x&\hat{\sigma}_z&\hspace{0.7cm}&\hat{\sigma}_x&\mathds{1}&\hat{\sigma}_y&\hat{\sigma}_y&\hat{\sigma}_z&\hat{\sigma}_z&\hspace{0.7cm}&\hat{\sigma}_x&\hat{\sigma}_y&\hat{\sigma}_y&\hat{\sigma}_x&\hat{\sigma}_x&\hat{\sigma}_z\\
\textstyle{}&\hspace{0.1cm}&&\mathds{1}&\mathds{1}&\hat{\sigma}_z&\mathds{1}&\mathds{1}&\hat{\sigma}_z&\hspace{0.7cm}&\mathds{1}&\hat{\sigma}_y&\hat{\sigma}_y&\mathds{1}&\hat{\sigma}_y&\hat{\sigma}_y&\hspace{0.7cm}&\hat{\sigma}_x&\mathds{1}&\hat{\sigma}_z&\hat{\sigma}_x&\mathds{1}&\hat{\sigma}_z&\hspace{0.7cm}&\hat{\sigma}_x&\hat{\sigma}_y&\hat{\sigma}_y&\hat{\sigma}_x&\hat{\sigma}_y&\hat{\sigma}_y\\
\textstyle{}&\hspace{0.1cm}&&\mathds{1}&\mathds{1}&\hat{\sigma}_z&\mathds{1}&\hat{\sigma}_z&\hat{\sigma}_z&\hspace{0.7cm}&\mathds{1}&\hat{\sigma}_y&\hat{\sigma}_y&\hat{\sigma}_z&\hat{\sigma}_x&\hat{\sigma}_z&\hspace{0.7cm}&\hat{\sigma}_x&\mathds{1}&\hat{\sigma}_z&\hat{\sigma}_x&\hat{\sigma}_z&\hat{\sigma}_z&\hspace{0.7cm}&\hat{\sigma}_x&\hat{\sigma}_y&\hat{\sigma}_y&\hat{\sigma}_y&\hat{\sigma}_x&\hat{\sigma}_y\\
\textstyle{01000}&\hspace{0.1cm}&&\mathds{1}&\mathds{1}&\hat{\sigma}_z&\hat{\sigma}_z&\mathds{1}&\hat{\sigma}_z&\hspace{0.7cm}&\mathds{1}&\hat{\sigma}_y&\hat{\sigma}_y&\hat{\sigma}_z&\hat{\sigma}_y&\hat{\sigma}_y&\hspace{0.7cm}&\hat{\sigma}_x&\mathds{1}&\hat{\sigma}_z&\hat{\sigma}_y&\mathds{1}&\hat{\sigma}_y&\hspace{0.7cm}&\hat{\sigma}_x&\hat{\sigma}_y&\hat{\sigma}_y&\hat{\sigma}_y&\hat{\sigma}_y&\hat{\sigma}_z\\
\textstyle{}&\hspace{0.1cm}&&\mathds{1}&\mathds{1}&\hat{\sigma}_z&\hat{\sigma}_z&\hat{\sigma}_z&\hat{\sigma}_z&\hspace{0.7cm}&\mathds{1}&\hat{\sigma}_y&\hat{\sigma}_z&\mathds{1}&\hat{\sigma}_x&\hat{\sigma}_y&\hspace{0.7cm}&\hat{\sigma}_x&\mathds{1}&\hat{\sigma}_z&\hat{\sigma}_y&\hat{\sigma}_z&\hat{\sigma}_y&\hspace{0.7cm}&\hat{\sigma}_x&\hat{\sigma}_y&\hat{\sigma}_z&\hat{\sigma}_x&\hat{\sigma}_x&\hat{\sigma}_y\\
\textstyle{}&\hspace{0.1cm}&&\mathds{1}&\hat{\sigma}_x&\mathds{1}&\mathds{1}&\hat{\sigma}_x&\mathds{1}&\hspace{0.7cm}&\mathds{1}&\hat{\sigma}_y&\hat{\sigma}_z&\mathds{1}&\hat{\sigma}_y&\hat{\sigma}_z&\hspace{0.7cm}&\hat{\sigma}_x&\hat{\sigma}_x&\mathds{1}&\hat{\sigma}_x&\hat{\sigma}_x&\mathds{1}&\hspace{0.7cm}&\hat{\sigma}_x&\hat{\sigma}_y&\hat{\sigma}_z&\hat{\sigma}_x&\hat{\sigma}_y&\hat{\sigma}_z\\
\textstyle{}&\hspace{0.1cm}&&\mathds{1}&\hat{\sigma}_x&\mathds{1}&\mathds{1}&\hat{\sigma}_x&\hat{\sigma}_x&\hspace{0.7cm}&\mathds{1}&\hat{\sigma}_y&\hat{\sigma}_z&\hat{\sigma}_z&\hat{\sigma}_x&\hat{\sigma}_y&\hspace{0.7cm}&\hat{\sigma}_x&\hat{\sigma}_x&\mathds{1}&\hat{\sigma}_x&\hat{\sigma}_x&\hat{\sigma}_x&\hspace{0.7cm}&\hat{\sigma}_x&\hat{\sigma}_y&\hat{\sigma}_z&\hat{\sigma}_y&\hat{\sigma}_x&\hat{\sigma}_z\\
\textstyle{01100}&\hspace{0.1cm}&&\mathds{1}&\hat{\sigma}_x&\mathds{1}&\hat{\sigma}_z&\hat{\sigma}_x&\mathds{1}&\hspace{0.7cm}&\mathds{1}&\hat{\sigma}_y&\hat{\sigma}_z&\hat{\sigma}_z&\hat{\sigma}_y&\hat{\sigma}_z&\hspace{0.7cm}&\hat{\sigma}_x&\hat{\sigma}_x&\mathds{1}&\hat{\sigma}_y&\hat{\sigma}_y&\mathds{1}&\hspace{0.7cm}&\hat{\sigma}_x&\hat{\sigma}_y&\hat{\sigma}_z&\hat{\sigma}_y&\hat{\sigma}_y&\hat{\sigma}_y\\
\textstyle{}&\hspace{0.1cm}&&\mathds{1}&\hat{\sigma}_x&\mathds{1}&\hat{\sigma}_z&\hat{\sigma}_x&\hat{\sigma}_x&\hspace{0.7cm}&\mathds{1}&\hat{\sigma}_z&\mathds{1}&\mathds{1}&\hat{\sigma}_z&\mathds{1}&\hspace{0.7cm}&\hat{\sigma}_x&\hat{\sigma}_x&\mathds{1}&\hat{\sigma}_y&\hat{\sigma}_y&\hat{\sigma}_x&\hspace{0.7cm}&\hat{\sigma}_x&\hat{\sigma}_z&\mathds{1}&\hat{\sigma}_x&\mathds{1}&\mathds{1}\\
\textstyle{}&\hspace{0.1cm}&&\mathds{1}&\hat{\sigma}_x&\hat{\sigma}_x&\mathds{1}&\hat{\sigma}_x&\mathds{1}&\hspace{0.7cm}&\mathds{1}&\hat{\sigma}_z&\hat{\sigma}_x&\mathds{1}&\hat{\sigma}_z&\hat{\sigma}_x&\hspace{0.7cm}&\hat{\sigma}_x&\hat{\sigma}_x&\hat{\sigma}_x&\hat{\sigma}_x&\hat{\sigma}_x&\mathds{1}&\hspace{0.7cm}&\hat{\sigma}_x&\hat{\sigma}_z&\mathds{1}&\hat{\sigma}_x&\mathds{1}&\hat{\sigma}_x\\
\textstyle{}&\hspace{0.1cm}&&\mathds{1}&\hat{\sigma}_x&\hat{\sigma}_x&\mathds{1}&\hat{\sigma}_x&\hat{\sigma}_x&\hspace{0.7cm}&\mathds{1}&\hat{\sigma}_z&\hat{\sigma}_y&\mathds{1}&\mathds{1}&\hat{\sigma}_y&\hspace{0.7cm}&\hat{\sigma}_x&\hat{\sigma}_x&\hat{\sigma}_x&\hat{\sigma}_x&\hat{\sigma}_x&\hat{\sigma}_x&\hspace{0.7cm}&\hat{\sigma}_x&\hat{\sigma}_z&\mathds{1}&\hat{\sigma}_x&\hat{\sigma}_z&\mathds{1}\\
\textstyle{10000}&\hspace{0.1cm}&&\mathds{1}&\hat{\sigma}_x&\hat{\sigma}_x&\hat{\sigma}_z&\hat{\sigma}_x&\mathds{1}&\hspace{0.7cm}&\mathds{1}&\hat{\sigma}_z&\hat{\sigma}_y&\mathds{1}&\hat{\sigma}_z&\hat{\sigma}_y&\hspace{0.7cm}&\hat{\sigma}_x&\hat{\sigma}_x&\hat{\sigma}_x&\hat{\sigma}_y&\hat{\sigma}_y&\mathds{1}&\hspace{0.7cm}&\hat{\sigma}_x&\hat{\sigma}_z&\mathds{1}&\hat{\sigma}_x&\hat{\sigma}_z&\hat{\sigma}_x\\
\textstyle{}&\hspace{0.1cm}&&\mathds{1}&\hat{\sigma}_x&\hat{\sigma}_x&\hat{\sigma}_z&\hat{\sigma}_x&\hat{\sigma}_x&\hspace{0.7cm}&\mathds{1}&\hat{\sigma}_z&\hat{\sigma}_y&\hat{\sigma}_z&\mathds{1}&\hat{\sigma}_y&\hspace{0.7cm}&\hat{\sigma}_x&\hat{\sigma}_x&\hat{\sigma}_x&\hat{\sigma}_y&\hat{\sigma}_y&\hat{\sigma}_x&\hspace{0.7cm}&\hat{\sigma}_x&\hat{\sigma}_z&\hat{\sigma}_x&\hat{\sigma}_x&\mathds{1}&\mathds{1}\\
\textstyle{}&\hspace{0.1cm}&&\mathds{1}&\hat{\sigma}_x&\hat{\sigma}_y&\mathds{1}&\hat{\sigma}_x&\hat{\sigma}_y&\hspace{0.7cm}&\mathds{1}&\hat{\sigma}_z&\hat{\sigma}_y&\hat{\sigma}_z&\hat{\sigma}_z&\hat{\sigma}_y&\hspace{0.7cm}&\hat{\sigma}_x&\hat{\sigma}_x&\hat{\sigma}_y&\hat{\sigma}_x&\hat{\sigma}_x&\hat{\sigma}_y&\hspace{0.7cm}&\hat{\sigma}_x&\hat{\sigma}_z&\hat{\sigma}_x&\hat{\sigma}_x&\mathds{1}&\hat{\sigma}_x\\
\textstyle{}&\hspace{0.1cm}&&\mathds{1}&\hat{\sigma}_x&\hat{\sigma}_y&\mathds{1}&\hat{\sigma}_y&\hat{\sigma}_z&\hspace{0.7cm}&\mathds{1}&\hat{\sigma}_z&\hat{\sigma}_z&\mathds{1}&\mathds{1}&\hat{\sigma}_z&\hspace{0.7cm}&\hat{\sigma}_x&\hat{\sigma}_x&\hat{\sigma}_y&\hat{\sigma}_x&\hat{\sigma}_y&\hat{\sigma}_z&\hspace{0.7cm}&\hat{\sigma}_x&\hat{\sigma}_z&\hat{\sigma}_x&\hat{\sigma}_x&\hat{\sigma}_z&\mathds{1}\\
\textstyle{10100}&\hspace{0.1cm}&&\mathds{1}&\hat{\sigma}_x&\hat{\sigma}_y&\hat{\sigma}_z&\hat{\sigma}_x&\hat{\sigma}_y&\hspace{0.7cm}&\mathds{1}&\hat{\sigma}_z&\hat{\sigma}_z&\mathds{1}&\hat{\sigma}_z&\hat{\sigma}_z&\hspace{0.7cm}&\hat{\sigma}_x&\hat{\sigma}_x&\hat{\sigma}_y&\hat{\sigma}_y&\hat{\sigma}_x&\hat{\sigma}_z&\hspace{0.7cm}&\hat{\sigma}_x&\hat{\sigma}_z&\hat{\sigma}_x&\hat{\sigma}_x&\hat{\sigma}_z&\hat{\sigma}_x\\
\textstyle{}&\hspace{0.1cm}&&\mathds{1}&\hat{\sigma}_x&\hat{\sigma}_y&\hat{\sigma}_z&\hat{\sigma}_y&\hat{\sigma}_z&\hspace{0.7cm}&\mathds{1}&\hat{\sigma}_z&\hat{\sigma}_z&\hat{\sigma}_z&\mathds{1}&\hat{\sigma}_z&\hspace{0.7cm}&\hat{\sigma}_x&\hat{\sigma}_x&\hat{\sigma}_y&\hat{\sigma}_y&\hat{\sigma}_y&\hat{\sigma}_y&\hspace{0.7cm}&\hat{\sigma}_x&\hat{\sigma}_z&\hat{\sigma}_y&\hat{\sigma}_x&\mathds{1}&\hat{\sigma}_y\\
\textstyle{}&\hspace{0.1cm}&&\mathds{1}&\hat{\sigma}_x&\hat{\sigma}_z&\mathds{1}&\hat{\sigma}_x&\hat{\sigma}_z&\hspace{0.7cm}&\mathds{1}&\hat{\sigma}_z&\hat{\sigma}_z&\hat{\sigma}_z&\hat{\sigma}_z&\hat{\sigma}_z&\hspace{0.7cm}&\hat{\sigma}_x&\hat{\sigma}_x&\hat{\sigma}_z&\hat{\sigma}_x&\hat{\sigma}_x&\hat{\sigma}_z&\hspace{0.7cm}&\hat{\sigma}_x&\hat{\sigma}_z&\hat{\sigma}_y&\hat{\sigma}_x&\hat{\sigma}_z&\hat{\sigma}_y\\
\textstyle{}&\hspace{0.1cm}&&\mathds{1}&\hat{\sigma}_x&\hat{\sigma}_z&\mathds{1}&\hat{\sigma}_y&\hat{\sigma}_y&\hspace{0.7cm}&\hat{\sigma}_x&\mathds{1}&\mathds{1}&\hat{\sigma}_x&\mathds{1}&\mathds{1}&\hspace{0.7cm}&\hat{\sigma}_x&\hat{\sigma}_x&\hat{\sigma}_z&\hat{\sigma}_x&\hat{\sigma}_y&\hat{\sigma}_y&\hspace{0.7cm}&\hat{\sigma}_x&\hat{\sigma}_z&\hat{\sigma}_y&\hat{\sigma}_y&\mathds{1}&\hat{\sigma}_z\\
\textstyle{11000}&\hspace{0.1cm}&&\mathds{1}&\hat{\sigma}_x&\hat{\sigma}_z&\hat{\sigma}_z&\hat{\sigma}_x&\hat{\sigma}_z&\hspace{0.7cm}&\hat{\sigma}_x&\mathds{1}&\mathds{1}&\hat{\sigma}_x&\mathds{1}&\hat{\sigma}_x&\hspace{0.7cm}&\hat{\sigma}_x&\hat{\sigma}_x&\hat{\sigma}_z&\hat{\sigma}_y&\hat{\sigma}_x&\hat{\sigma}_y&\hspace{0.7cm}&\hat{\sigma}_x&\hat{\sigma}_z&\hat{\sigma}_y&\hat{\sigma}_y&\hat{\sigma}_z&\hat{\sigma}_z\\
\textstyle{}&\hspace{0.1cm}&&\mathds{1}&\hat{\sigma}_x&\hat{\sigma}_z&\hat{\sigma}_z&\hat{\sigma}_y&\hat{\sigma}_y&\hspace{0.7cm}&\hat{\sigma}_x&\mathds{1}&\mathds{1}&\hat{\sigma}_x&\hat{\sigma}_z&\mathds{1}&\hspace{0.7cm}&\hat{\sigma}_x&\hat{\sigma}_x&\hat{\sigma}_z&\hat{\sigma}_y&\hat{\sigma}_y&\hat{\sigma}_z&\hspace{0.7cm}&\hat{\sigma}_x&\hat{\sigma}_z&\hat{\sigma}_z&\hat{\sigma}_x&\mathds{1}&\hat{\sigma}_z\\
\textstyle{}&\hspace{0.1cm}&&\mathds{1}&\hat{\sigma}_y&\mathds{1}&\mathds{1}&\hat{\sigma}_y&\mathds{1}&\hspace{0.7cm}&\hat{\sigma}_x&\mathds{1}&\mathds{1}&\hat{\sigma}_x&\hat{\sigma}_z&\hat{\sigma}_x&\hspace{0.7cm}&\hat{\sigma}_x&\hat{\sigma}_y&\mathds{1}&\hat{\sigma}_x&\hat{\sigma}_y&\mathds{1}&\hspace{0.7cm}&\hat{\sigma}_x&\hat{\sigma}_z&\hat{\sigma}_z&\hat{\sigma}_x&\hat{\sigma}_z&\hat{\sigma}_z\\
\textstyle{}&\hspace{0.1cm}&&\mathds{1}&\hat{\sigma}_y&\mathds{1}&\mathds{1}&\hat{\sigma}_y&\hat{\sigma}_x&\hspace{0.7cm}&\hat{\sigma}_x&\mathds{1}&\hat{\sigma}_x&\hat{\sigma}_x&\mathds{1}&\mathds{1}&\hspace{0.7cm}&\hat{\sigma}_x&\hat{\sigma}_y&\mathds{1}&\hat{\sigma}_x&\hat{\sigma}_y&\hat{\sigma}_x&\hspace{0.7cm}&\hat{\sigma}_x&\hat{\sigma}_z&\hat{\sigma}_z&\hat{\sigma}_y&\mathds{1}&\hat{\sigma}_y\\
\textstyle{11100}&\hspace{0.1cm}&&\mathds{1}&\hat{\sigma}_y&\mathds{1}&\hat{\sigma}_z&\hat{\sigma}_y&\mathds{1}&\hspace{0.7cm}&\hat{\sigma}_x&\mathds{1}&\hat{\sigma}_x&\hat{\sigma}_x&\mathds{1}&\hat{\sigma}_x&\hspace{0.7cm}&\hat{\sigma}_x&\hat{\sigma}_y&\mathds{1}&\hat{\sigma}_y&\hat{\sigma}_x&\mathds{1}&\hspace{0.7cm}&\hat{\sigma}_x&\hat{\sigma}_z&\hat{\sigma}_z&\hat{\sigma}_y&\hat{\sigma}_z&\hat{\sigma}_y\\
\end{array}
\end{math}
\caption{Elements 00000000 to 01111100}
\label{tab:toffoli1}
\end{table*}

\begin{table*}
\begin{math}
\begin{array}{llllllllllllllllllllllllllllll}
&&&\multicolumn{6}{c}{100}&&\multicolumn{6}{c}{101}&&\multicolumn{6}{c}{110}&&\multicolumn{6}{c}{111}\\\\\textstyle{00000}&\hspace{0.1cm}&&\hat{\sigma}_y&\mathds{1}&\mathds{1}&\hat{\sigma}_y&\mathds{1}&\mathds{1}&\hspace{0.7cm}&\hat{\sigma}_y&\hat{\sigma}_x&\hat{\sigma}_z&\hat{\sigma}_x&\hat{\sigma}_y&\hat{\sigma}_z&\hspace{0.7cm}&\hat{\sigma}_y&\hat{\sigma}_z&\hat{\sigma}_y&\hat{\sigma}_y&\mathds{1}&\hat{\sigma}_y&\hspace{0.7cm}&\hat{\sigma}_z&\hat{\sigma}_x&\hat{\sigma}_z&\mathds{1}&\hat{\sigma}_y&\hat{\sigma}_y\\
\textstyle{}&\hspace{0.1cm}&&\hat{\sigma}_y&\mathds{1}&\mathds{1}&\hat{\sigma}_y&\mathds{1}&\hat{\sigma}_x&\hspace{0.7cm}&\hat{\sigma}_y&\hat{\sigma}_x&\hat{\sigma}_z&\hat{\sigma}_y&\hat{\sigma}_x&\hat{\sigma}_z&\hspace{0.7cm}&\hat{\sigma}_y&\hat{\sigma}_z&\hat{\sigma}_y&\hat{\sigma}_y&\hat{\sigma}_z&\hat{\sigma}_y&\hspace{0.7cm}&\hat{\sigma}_z&\hat{\sigma}_x&\hat{\sigma}_z&\hat{\sigma}_z&\hat{\sigma}_x&\hat{\sigma}_z\\
\textstyle{}&\hspace{0.1cm}&&\hat{\sigma}_y&\mathds{1}&\mathds{1}&\hat{\sigma}_y&\hat{\sigma}_z&\mathds{1}&\hspace{0.7cm}&\hat{\sigma}_y&\hat{\sigma}_x&\hat{\sigma}_z&\hat{\sigma}_y&\hat{\sigma}_y&\hat{\sigma}_y&\hspace{0.7cm}&\hat{\sigma}_y&\hat{\sigma}_z&\hat{\sigma}_z&\hat{\sigma}_x&\mathds{1}&\hat{\sigma}_y&\hspace{0.7cm}&\hat{\sigma}_z&\hat{\sigma}_x&\hat{\sigma}_z&\hat{\sigma}_z&\hat{\sigma}_y&\hat{\sigma}_y\\
\textstyle{}&\hspace{0.1cm}&&\hat{\sigma}_y&\mathds{1}&\mathds{1}&\hat{\sigma}_y&\hat{\sigma}_z&\hat{\sigma}_x&\hspace{0.7cm}&\hat{\sigma}_y&\hat{\sigma}_y&\mathds{1}&\hat{\sigma}_x&\hat{\sigma}_x&\mathds{1}&\hspace{0.7cm}&\hat{\sigma}_y&\hat{\sigma}_z&\hat{\sigma}_z&\hat{\sigma}_x&\hat{\sigma}_z&\hat{\sigma}_y&\hspace{0.7cm}&\hat{\sigma}_z&\hat{\sigma}_y&\mathds{1}&\mathds{1}&\hat{\sigma}_y&\mathds{1}\\
\textstyle{00100}&\hspace{0.1cm}&&\hat{\sigma}_y&\mathds{1}&\hat{\sigma}_x&\hat{\sigma}_y&\mathds{1}&\mathds{1}&\hspace{0.7cm}&\hat{\sigma}_y&\hat{\sigma}_y&\mathds{1}&\hat{\sigma}_x&\hat{\sigma}_x&\hat{\sigma}_x&\hspace{0.7cm}&\hat{\sigma}_y&\hat{\sigma}_z&\hat{\sigma}_z&\hat{\sigma}_y&\mathds{1}&\hat{\sigma}_z&\hspace{0.7cm}&\hat{\sigma}_z&\hat{\sigma}_y&\mathds{1}&\mathds{1}&\hat{\sigma}_y&\hat{\sigma}_x\\
\textstyle{}&\hspace{0.1cm}&&\hat{\sigma}_y&\mathds{1}&\hat{\sigma}_x&\hat{\sigma}_y&\mathds{1}&\hat{\sigma}_x&\hspace{0.7cm}&\hat{\sigma}_y&\hat{\sigma}_y&\mathds{1}&\hat{\sigma}_y&\hat{\sigma}_y&\mathds{1}&\hspace{0.7cm}&\hat{\sigma}_y&\hat{\sigma}_z&\hat{\sigma}_z&\hat{\sigma}_y&\hat{\sigma}_z&\hat{\sigma}_z&\hspace{0.7cm}&\hat{\sigma}_z&\hat{\sigma}_y&\mathds{1}&\hat{\sigma}_z&\hat{\sigma}_y&\mathds{1}\\
\textstyle{}&\hspace{0.1cm}&&\hat{\sigma}_y&\mathds{1}&\hat{\sigma}_x&\hat{\sigma}_y&\hat{\sigma}_z&\mathds{1}&\hspace{0.7cm}&\hat{\sigma}_y&\hat{\sigma}_y&\mathds{1}&\hat{\sigma}_y&\hat{\sigma}_y&\hat{\sigma}_x&\hspace{0.7cm}&\hat{\sigma}_z&\mathds{1}&\mathds{1}&\hat{\sigma}_z&\mathds{1}&\mathds{1}&\hspace{0.7cm}&\hat{\sigma}_z&\hat{\sigma}_y&\mathds{1}&\hat{\sigma}_z&\hat{\sigma}_y&\hat{\sigma}_x\\
\textstyle{}&\hspace{0.1cm}&&\hat{\sigma}_y&\mathds{1}&\hat{\sigma}_x&\hat{\sigma}_y&\hat{\sigma}_z&\hat{\sigma}_x&\hspace{0.7cm}&\hat{\sigma}_y&\hat{\sigma}_y&\hat{\sigma}_x&\hat{\sigma}_x&\hat{\sigma}_x&\mathds{1}&\hspace{0.7cm}&\hat{\sigma}_z&\mathds{1}&\hat{\sigma}_x&\hat{\sigma}_z&\mathds{1}&\hat{\sigma}_x&\hspace{0.7cm}&\hat{\sigma}_z&\hat{\sigma}_y&\hat{\sigma}_x&\mathds{1}&\hat{\sigma}_y&\mathds{1}\\
\textstyle{01000}&\hspace{0.1cm}&&\hat{\sigma}_y&\mathds{1}&\hat{\sigma}_y&\hat{\sigma}_x&\mathds{1}&\hat{\sigma}_z&\hspace{0.7cm}&\hat{\sigma}_y&\hat{\sigma}_y&\hat{\sigma}_x&\hat{\sigma}_x&\hat{\sigma}_x&\hat{\sigma}_x&\hspace{0.7cm}&\hat{\sigma}_z&\mathds{1}&\hat{\sigma}_y&\mathds{1}&\mathds{1}&\hat{\sigma}_y&\hspace{0.7cm}&\hat{\sigma}_z&\hat{\sigma}_y&\hat{\sigma}_x&\mathds{1}&\hat{\sigma}_y&\hat{\sigma}_x\\
\textstyle{}&\hspace{0.1cm}&&\hat{\sigma}_y&\mathds{1}&\hat{\sigma}_y&\hat{\sigma}_x&\hat{\sigma}_z&\hat{\sigma}_z&\hspace{0.7cm}&\hat{\sigma}_y&\hat{\sigma}_y&\hat{\sigma}_x&\hat{\sigma}_y&\hat{\sigma}_y&\mathds{1}&\hspace{0.7cm}&\hat{\sigma}_z&\mathds{1}&\hat{\sigma}_y&\mathds{1}&\hat{\sigma}_z&\hat{\sigma}_y&\hspace{0.7cm}&\hat{\sigma}_z&\hat{\sigma}_y&\hat{\sigma}_x&\hat{\sigma}_z&\hat{\sigma}_y&\mathds{1}\\
\textstyle{}&\hspace{0.1cm}&&\hat{\sigma}_y&\mathds{1}&\hat{\sigma}_y&\hat{\sigma}_y&\mathds{1}&\hat{\sigma}_y&\hspace{0.7cm}&\hat{\sigma}_y&\hat{\sigma}_y&\hat{\sigma}_x&\hat{\sigma}_y&\hat{\sigma}_y&\hat{\sigma}_x&\hspace{0.7cm}&\hat{\sigma}_z&\mathds{1}&\hat{\sigma}_y&\hat{\sigma}_z&\mathds{1}&\hat{\sigma}_y&\hspace{0.7cm}&\hat{\sigma}_z&\hat{\sigma}_y&\hat{\sigma}_x&\hat{\sigma}_z&\hat{\sigma}_y&\hat{\sigma}_x\\
\textstyle{}&\hspace{0.1cm}&&\hat{\sigma}_y&\mathds{1}&\hat{\sigma}_y&\hat{\sigma}_y&\hat{\sigma}_z&\hat{\sigma}_y&\hspace{0.7cm}&\hat{\sigma}_y&\hat{\sigma}_y&\hat{\sigma}_y&\hat{\sigma}_x&\hat{\sigma}_x&\hat{\sigma}_y&\hspace{0.7cm}&\hat{\sigma}_z&\mathds{1}&\hat{\sigma}_y&\hat{\sigma}_z&\hat{\sigma}_z&\hat{\sigma}_y&\hspace{0.7cm}&\hat{\sigma}_z&\hat{\sigma}_y&\hat{\sigma}_y&\mathds{1}&\hat{\sigma}_x&\hat{\sigma}_z\\
\textstyle{01100}&\hspace{0.1cm}&&\hat{\sigma}_y&\mathds{1}&\hat{\sigma}_z&\hat{\sigma}_x&\mathds{1}&\hat{\sigma}_y&\hspace{0.7cm}&\hat{\sigma}_y&\hat{\sigma}_y&\hat{\sigma}_y&\hat{\sigma}_x&\hat{\sigma}_y&\hat{\sigma}_z&\hspace{0.7cm}&\hat{\sigma}_z&\mathds{1}&\hat{\sigma}_z&\mathds{1}&\mathds{1}&\hat{\sigma}_z&\hspace{0.7cm}&\hat{\sigma}_z&\hat{\sigma}_y&\hat{\sigma}_y&\mathds{1}&\hat{\sigma}_y&\hat{\sigma}_y\\
\textstyle{}&\hspace{0.1cm}&&\hat{\sigma}_y&\mathds{1}&\hat{\sigma}_z&\hat{\sigma}_x&\hat{\sigma}_z&\hat{\sigma}_y&\hspace{0.7cm}&\hat{\sigma}_y&\hat{\sigma}_y&\hat{\sigma}_y&\hat{\sigma}_y&\hat{\sigma}_x&\hat{\sigma}_z&\hspace{0.7cm}&\hat{\sigma}_z&\mathds{1}&\hat{\sigma}_z&\mathds{1}&\hat{\sigma}_z&\hat{\sigma}_z&\hspace{0.7cm}&\hat{\sigma}_z&\hat{\sigma}_y&\hat{\sigma}_y&\hat{\sigma}_z&\hat{\sigma}_x&\hat{\sigma}_z\\
\textstyle{}&\hspace{0.1cm}&&\hat{\sigma}_y&\mathds{1}&\hat{\sigma}_z&\hat{\sigma}_y&\mathds{1}&\hat{\sigma}_z&\hspace{0.7cm}&\hat{\sigma}_y&\hat{\sigma}_y&\hat{\sigma}_y&\hat{\sigma}_y&\hat{\sigma}_y&\hat{\sigma}_y&\hspace{0.7cm}&\hat{\sigma}_z&\mathds{1}&\hat{\sigma}_z&\hat{\sigma}_z&\mathds{1}&\hat{\sigma}_z&\hspace{0.7cm}&\hat{\sigma}_z&\hat{\sigma}_y&\hat{\sigma}_y&\hat{\sigma}_z&\hat{\sigma}_y&\hat{\sigma}_y\\
\textstyle{}&\hspace{0.1cm}&&\hat{\sigma}_y&\mathds{1}&\hat{\sigma}_z&\hat{\sigma}_y&\hat{\sigma}_z&\hat{\sigma}_z&\hspace{0.7cm}&\hat{\sigma}_y&\hat{\sigma}_y&\hat{\sigma}_z&\hat{\sigma}_x&\hat{\sigma}_x&\hat{\sigma}_z&\hspace{0.7cm}&\hat{\sigma}_z&\mathds{1}&\hat{\sigma}_z&\hat{\sigma}_z&\hat{\sigma}_z&\hat{\sigma}_z&\hspace{0.7cm}&\hat{\sigma}_z&\hat{\sigma}_y&\hat{\sigma}_z&\mathds{1}&\hat{\sigma}_x&\hat{\sigma}_y\\
\textstyle{10000}&\hspace{0.1cm}&&\hat{\sigma}_y&\hat{\sigma}_x&\mathds{1}&\hat{\sigma}_x&\hat{\sigma}_y&\mathds{1}&\hspace{0.7cm}&\hat{\sigma}_y&\hat{\sigma}_y&\hat{\sigma}_z&\hat{\sigma}_x&\hat{\sigma}_y&\hat{\sigma}_y&\hspace{0.7cm}&\hat{\sigma}_z&\hat{\sigma}_x&\mathds{1}&\mathds{1}&\hat{\sigma}_x&\mathds{1}&\hspace{0.7cm}&\hat{\sigma}_z&\hat{\sigma}_y&\hat{\sigma}_z&\mathds{1}&\hat{\sigma}_y&\hat{\sigma}_z\\
\textstyle{}&\hspace{0.1cm}&&\hat{\sigma}_y&\hat{\sigma}_x&\mathds{1}&\hat{\sigma}_x&\hat{\sigma}_y&\hat{\sigma}_x&\hspace{0.7cm}&\hat{\sigma}_y&\hat{\sigma}_y&\hat{\sigma}_z&\hat{\sigma}_y&\hat{\sigma}_x&\hat{\sigma}_y&\hspace{0.7cm}&\hat{\sigma}_z&\hat{\sigma}_x&\mathds{1}&\mathds{1}&\hat{\sigma}_x&\hat{\sigma}_x&\hspace{0.7cm}&\hat{\sigma}_z&\hat{\sigma}_y&\hat{\sigma}_z&\hat{\sigma}_z&\hat{\sigma}_x&\hat{\sigma}_y\\
\textstyle{}&\hspace{0.1cm}&&\hat{\sigma}_y&\hat{\sigma}_x&\mathds{1}&\hat{\sigma}_y&\hat{\sigma}_x&\mathds{1}&\hspace{0.7cm}&\hat{\sigma}_y&\hat{\sigma}_y&\hat{\sigma}_z&\hat{\sigma}_y&\hat{\sigma}_y&\hat{\sigma}_z&\hspace{0.7cm}&\hat{\sigma}_z&\hat{\sigma}_x&\mathds{1}&\hat{\sigma}_z&\hat{\sigma}_x&\mathds{1}&\hspace{0.7cm}&\hat{\sigma}_z&\hat{\sigma}_y&\hat{\sigma}_z&\hat{\sigma}_z&\hat{\sigma}_y&\hat{\sigma}_z\\
\textstyle{}&\hspace{0.1cm}&&\hat{\sigma}_y&\hat{\sigma}_x&\mathds{1}&\hat{\sigma}_y&\hat{\sigma}_x&\hat{\sigma}_x&\hspace{0.7cm}&\hat{\sigma}_y&\hat{\sigma}_z&\mathds{1}&\hat{\sigma}_y&\mathds{1}&\mathds{1}&\hspace{0.7cm}&\hat{\sigma}_z&\hat{\sigma}_x&\mathds{1}&\hat{\sigma}_z&\hat{\sigma}_x&\hat{\sigma}_x&\hspace{0.7cm}&\hat{\sigma}_z&\hat{\sigma}_z&\mathds{1}&\hat{\sigma}_z&\hat{\sigma}_z&\mathds{1}\\
\textstyle{10100}&\hspace{0.1cm}&&\hat{\sigma}_y&\hat{\sigma}_x&\hat{\sigma}_x&\hat{\sigma}_x&\hat{\sigma}_y&\mathds{1}&\hspace{0.7cm}&\hat{\sigma}_y&\hat{\sigma}_z&\mathds{1}&\hat{\sigma}_y&\mathds{1}&\hat{\sigma}_x&\hspace{0.7cm}&\hat{\sigma}_z&\hat{\sigma}_x&\hat{\sigma}_x&\mathds{1}&\hat{\sigma}_x&\mathds{1}&\hspace{0.7cm}&\hat{\sigma}_z&\hat{\sigma}_z&\hat{\sigma}_x&\hat{\sigma}_z&\hat{\sigma}_z&\hat{\sigma}_x\\
\textstyle{}&\hspace{0.1cm}&&\hat{\sigma}_y&\hat{\sigma}_x&\hat{\sigma}_x&\hat{\sigma}_x&\hat{\sigma}_y&\hat{\sigma}_x&\hspace{0.7cm}&\hat{\sigma}_y&\hat{\sigma}_z&\mathds{1}&\hat{\sigma}_y&\hat{\sigma}_z&\mathds{1}&\hspace{0.7cm}&\hat{\sigma}_z&\hat{\sigma}_x&\hat{\sigma}_x&\mathds{1}&\hat{\sigma}_x&\hat{\sigma}_x&\hspace{0.7cm}&\hat{\sigma}_z&\hat{\sigma}_z&\hat{\sigma}_y&\mathds{1}&\mathds{1}&\hat{\sigma}_y\\
\textstyle{}&\hspace{0.1cm}&&\hat{\sigma}_y&\hat{\sigma}_x&\hat{\sigma}_x&\hat{\sigma}_y&\hat{\sigma}_x&\mathds{1}&\hspace{0.7cm}&\hat{\sigma}_y&\hat{\sigma}_z&\mathds{1}&\hat{\sigma}_y&\hat{\sigma}_z&\hat{\sigma}_x&\hspace{0.7cm}&\hat{\sigma}_z&\hat{\sigma}_x&\hat{\sigma}_x&\hat{\sigma}_z&\hat{\sigma}_x&\mathds{1}&\hspace{0.7cm}&\hat{\sigma}_z&\hat{\sigma}_z&\hat{\sigma}_y&\mathds{1}&\hat{\sigma}_z&\hat{\sigma}_y\\
\textstyle{}&\hspace{0.1cm}&&\hat{\sigma}_y&\hat{\sigma}_x&\hat{\sigma}_x&\hat{\sigma}_y&\hat{\sigma}_x&\hat{\sigma}_x&\hspace{0.7cm}&\hat{\sigma}_y&\hat{\sigma}_z&\hat{\sigma}_x&\hat{\sigma}_y&\mathds{1}&\mathds{1}&\hspace{0.7cm}&\hat{\sigma}_z&\hat{\sigma}_x&\hat{\sigma}_x&\hat{\sigma}_z&\hat{\sigma}_x&\hat{\sigma}_x&\hspace{0.7cm}&\hat{\sigma}_z&\hat{\sigma}_z&\hat{\sigma}_y&\hat{\sigma}_z&\mathds{1}&\hat{\sigma}_y\\
\textstyle{11000}&\hspace{0.1cm}&&\hat{\sigma}_y&\hat{\sigma}_x&\hat{\sigma}_y&\hat{\sigma}_x&\hat{\sigma}_x&\hat{\sigma}_z&\hspace{0.7cm}&\hat{\sigma}_y&\hat{\sigma}_z&\hat{\sigma}_x&\hat{\sigma}_y&\mathds{1}&\hat{\sigma}_x&\hspace{0.7cm}&\hat{\sigma}_z&\hat{\sigma}_x&\hat{\sigma}_y&\mathds{1}&\hat{\sigma}_x&\hat{\sigma}_y&\hspace{0.7cm}&\hat{\sigma}_z&\hat{\sigma}_z&\hat{\sigma}_y&\hat{\sigma}_z&\hat{\sigma}_z&\hat{\sigma}_y\\
\textstyle{}&\hspace{0.1cm}&&\hat{\sigma}_y&\hat{\sigma}_x&\hat{\sigma}_y&\hat{\sigma}_x&\hat{\sigma}_y&\hat{\sigma}_y&\hspace{0.7cm}&\hat{\sigma}_y&\hat{\sigma}_z&\hat{\sigma}_x&\hat{\sigma}_y&\hat{\sigma}_z&\mathds{1}&\hspace{0.7cm}&\hat{\sigma}_z&\hat{\sigma}_x&\hat{\sigma}_y&\mathds{1}&\hat{\sigma}_y&\hat{\sigma}_z&\hspace{0.7cm}&\hat{\sigma}_z&\hat{\sigma}_z&\hat{\sigma}_z&\mathds{1}&\mathds{1}&\hat{\sigma}_z\\
\textstyle{}&\hspace{0.1cm}&&\hat{\sigma}_y&\hat{\sigma}_x&\hat{\sigma}_y&\hat{\sigma}_y&\hat{\sigma}_x&\hat{\sigma}_y&\hspace{0.7cm}&\hat{\sigma}_y&\hat{\sigma}_z&\hat{\sigma}_x&\hat{\sigma}_y&\hat{\sigma}_z&\hat{\sigma}_x&\hspace{0.7cm}&\hat{\sigma}_z&\hat{\sigma}_x&\hat{\sigma}_y&\hat{\sigma}_z&\hat{\sigma}_x&\hat{\sigma}_y&\hspace{0.7cm}&\hat{\sigma}_z&\hat{\sigma}_z&\hat{\sigma}_z&\mathds{1}&\hat{\sigma}_z&\hat{\sigma}_z\\
\textstyle{}&\hspace{0.1cm}&&\hat{\sigma}_y&\hat{\sigma}_x&\hat{\sigma}_y&\hat{\sigma}_y&\hat{\sigma}_y&\hat{\sigma}_z&\hspace{0.7cm}&\hat{\sigma}_y&\hat{\sigma}_z&\hat{\sigma}_y&\hat{\sigma}_x&\mathds{1}&\hat{\sigma}_z&\hspace{0.7cm}&\hat{\sigma}_z&\hat{\sigma}_x&\hat{\sigma}_y&\hat{\sigma}_z&\hat{\sigma}_y&\hat{\sigma}_z&\hspace{0.7cm}&\hat{\sigma}_z&\hat{\sigma}_z&\hat{\sigma}_z&\hat{\sigma}_z&\mathds{1}&\hat{\sigma}_z\\
\textstyle{11100}&\hspace{0.1cm}&&\hat{\sigma}_y&\hat{\sigma}_x&\hat{\sigma}_z&\hat{\sigma}_x&\hat{\sigma}_x&\hat{\sigma}_y&\hspace{0.7cm}&\hat{\sigma}_y&\hat{\sigma}_z&\hat{\sigma}_y&\hat{\sigma}_x&\hat{\sigma}_z&\hat{\sigma}_z&\hspace{0.7cm}&\hat{\sigma}_z&\hat{\sigma}_x&\hat{\sigma}_z&\mathds{1}&\hat{\sigma}_x&\hat{\sigma}_z&\hspace{0.7cm}&\hat{\sigma}_z&\hat{\sigma}_z&\hat{\sigma}_z&\hat{\sigma}_z&\hat{\sigma}_z&\hat{\sigma}_z\\
\end{array}
\end{math}
\caption{Elements 10000000 to 11111100}
\label{tab:toffoli2}
\end{table*}

\bibliographystyle{apsrev4-1}
\bibliography{Q:/USERS/Lars/refdb/QudevRefDB}
